\documentclass[10pt]{iopart}
\usepackage{iopams} 
\usepackage{graphicx, bm, color}
\usepackage{dblfloatfix}
\usepackage{upgreek}
\usepackage{epstopdf}
\usepackage{color}
\begin{document}

\title{Performance-defining properties of Nb$_3$Sn coating in SRF cavities}

\author{Y. Trenikhina}
\address{Fermi National Accelerator Laboratory, Batavia, IL 60510, USA.}
\ead{yuliatr@fnal.gov}
\vspace{10pt}

\author{S. Posen}
\address{Fermi National Accelerator Laboratory, Batavia, IL 60510, USA.}
\ead{sposen@fnal.gov}

\author{A. Romanenko}
\address{Fermi National Accelerator Laboratory, Batavia, IL 60510, USA.}
\ead{aroman@fnal.gov}

\author{M. Sardela}
\address{Materials Science and Engineering Department, University of Illinois, Urbana IL 61801, United States}
\ead{sardela@illinois.edu}

\author{J.-M. Zuo}
\address{Materials Science and Engineering Department, University of Illinois, Urbana IL 61801, United States}
\ead{jianzuo@illinois.edu}

\author{D. L. Hall}
\address{Cornell Laboratory for Accelerator-Based Sciences and Education, Ithaca, NY, 14853, USA}
\ead{dlh269@cornell.edu}

\author{M. Liepe}
\address{Cornell Laboratory for Accelerator-Based Sciences and Education, Ithaca, NY, 14853, USA}
\ead{mul2@cornell.edu}
\vspace{10pt}

\begin{abstract}
Nb$_3$Sn has great potential to become the material of choice for fabrication of SRF cavities.  The higher critical temperature of Nb$_3$Sn potentially allows for an increased operational temperature in SRF cavities, which could enable tremendous simplification of cryogenic system, leading to significant cost reduction.  We present extended characterization of a Nb$_3$Sn coated Nb cavity prepared at Cornell University.  Using combination of thermometry during cavity RF measurements, and structural and analytical characterization of the cavity cutouts, we discover Nb$_3$Sn coating flaws responsible for the poor cavity performance.  Our results clearly show degraded material quality in the cavity cutouts which exhibit significant heating during the RF testing.  Results of extended comparison of cavity cutouts with different dissipation profiles not only point out the cause of significant Q-slope but also establish figures of merit for material qualities in relation to the quality of SRF performance.
\end{abstract}

%
%
%
%
\ioptwocol

\section{Introduction}

Superconducting radio frequency (SRF) cavities are the technology of choice for the most modern particle accelerators \cite{Hasan_book2,Padamsee_Ann_Rev_Nucl_2014}.  Traditionally SRF cavities are made of bulk niobium, which has superconducting critical temperature of T$_\mathrm{c}=9.25$~K.  Growing demand for higher SRF cavity performance at higher accelerating gradients brings traditional niobium to it's theoretical limits.  Nb$_3$Sn alloy is one of the primary alternative materials for SRF cavity production.  Being a type II superconductor with maximum T$_\mathrm{c}$ of 18K and superheating field of 400 mT \cite{Transtrum}, Nb$_3$Sn offers a much broader parameter space for the new SRF performance records. 

An optimized physical vapor deposition method adopted at Cornell University for production of Nb$_3$Sn coating on Nb cavities has shown the great potential of Nb$_3$Sn for SRF applications \cite{Posen-APL-2015, Posen-Phys.Rev-2014}.  Despite encouraging development of Nb3Sn technology for the SRF field, systematic understanding is absent for the cause of quench fields which has been observed in the 14-17 MV/m range and for the Q-slope observed in some coated cavities. This Q-slope is characterized by a decrease of the quality factor (increase in surface resistance) with accelerating gradient, which is associated with increased dissipation due to heating.  Significant heating of the cavity wall may cause loss of superconductivity, which leads to quench.

The major challenge is to understand why the expected theoretical level of Nb$_3$Sn performance is still out of reach.  In order to understand and overcome cavity performance limitations, fundamental intrinsic properties of Nb$_3$Sn coating need to be evaluated.  While construction of Fermilab's Nb$_3$Sn coating facility was in progress \cite{Posen-SRF-2015}, we initiated extended characterization of the first Nb$_3$Sn coated cavity from Cornell University.  Though subsequent cavities coated at Cornell showed much higher quality factors \cite{Posen-2015-APL}, performance of this 1.3 GHz single cell Nb$_3$Sn coated cavity was severely limited (Q$_0$ in the order of 10$^9$) by strong heating concentrated in one of the half cells.  Finding connection between Nb$_3$Sn material properties and well-characterized RF performance is our pathway to understand what Nb$_3$Sn features are responsible for the cavity performance degradation and how these features had formed during the deposition process.  Cavity with prominent Q-slope gives us a perfect opportunity to study performance-limiting defects in Nb$_3$Sn.

This paper discusses structural and analytical characterization of Nb$_3$Sn coating in cavity cutouts with different dissipation characteristics.  We use X-Ray Diffraction (XRD), Scanning Electron Microscopy (SEM), Transmission Electron Microscopy (TEM), Scanning Transmission Electron Microscopy (STEM) and Energy Dispersive Spectroscopy (EDS).  Use of a temperature mapping system during the RF-measurements of  Nb$_3$Sn-coated Nb cavity enables us to directly relate properties of the coating to the local dissipation profiles.  The use of multiple material characterization techniques provides complete insight into composition, homogeneity and structure of Nb$_3$Sn coating for SRF application.

\section{Experimental Methods}

\subsection{Identification of the cutout regions}

\begin{figure}
\includegraphics[width=0.5\textwidth]{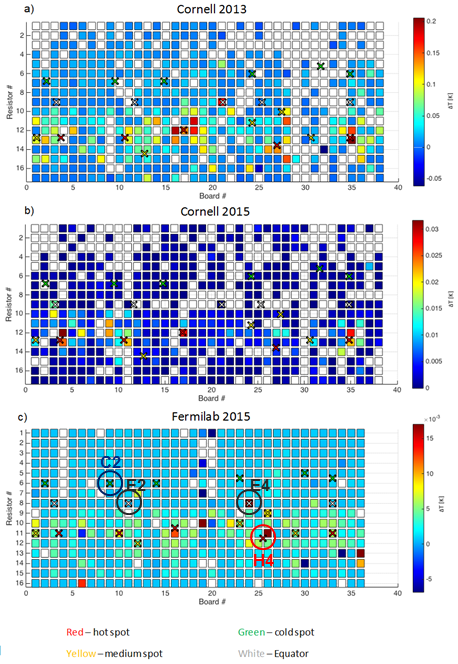}
\centering
\caption{Temperature maps showing excessive heating in one half cell (regions below resistor 9 on the Cornell T-map and below resistor 8 on the Fermilab T-map). First measurements were taken at Cornell in 2013 (a) then new studies were performed in 2015 at Cornell (b) and FNAL (c).}
\label{T-map}
\end{figure}

1.3 GHz single cell Nb cavity coated with Nb$_3$Sn (ERL1-5) at Cornell University \cite{Posen-Thesis} had been RF characterized at Cornell in 2013 and at Fermilab in 2015 (Fig.$\ref{T-map}$).  The first cavity coated under Cornell's program \cite{Posen-SRF-2015-049}, ERL1-5  demonstrated significantly degraded performance compared to the cavities which had been Nb$_3$Sn-coated afterwards.  Temperature map shows significant heating which is mostly present in one half-cell.  Visual appearance of two half-cells seems to be different.  Looking into the surfaces of the cavity visible from the beamtubes, one half-cell appeared a matte gray as expected, while the other appeared shiny.  Attempts to re-coat the same cavity resulted in the similar appearance of half-cells and similar quality of performance \cite{Posen-Thesis}.

\begin{figure}
\includegraphics[width=0.34\textwidth]{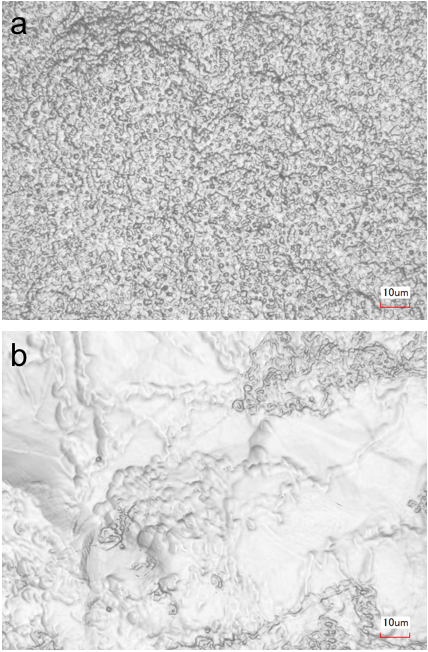}
\centering
\caption{Images of (a) cold and (b) hot cutouts from optical microscope.}
\label{optical}
\end{figure}

Temperature maps indicate that the half-cell with the unusually shiny appearance has large regions with far a higher surface resistance than is observed in the other half cell.  The intention of repeated RF characterization at Fermilab was to map out well the surface resistance of the cavity, so that the ``hot'' and ``cold'' regions could be identified with a high degree of confidence.  T-map was measured at 9 MV/m at 4.2 K at Fermilab.   Coupons located at these hot and cold spots were subsequently cut out from the cavity.  Cutouts were in a 1cm diameter, circular shape.  Two cutouts were depicted for the analysis.  Spot marked as ``H4" on the temperature map, was taken from the shiny half-cell as representative of strongly dissipating region (referred as hot spot).  ``C2" on the temperature map indicates the location of non-dissipating cold spot taken from the matte half-cell (Fig.$\ref{T-map}$c).  Fig.$\ref{optical}$a and b shows optical images of cold cutout (C4) and hot cutout (H1), respectively.

Direct comparison of the original H4 and C2 cutouts allows us to relate differences in the coating quality to the dissipation profiles.  The ultimate goal of our work is not only to find out coating qualities which lead to the unsatisfactory performance but also to understand why poor coating takes place. 

In addition to the hot and cold cutouts, two cutouts taken from the equator region of the cavity (referred as equator cutouts, E2 and E4) were characterized.

\subsection{Characterization Techniques}

Cross sectional TEM samples were prepared from Nb$_3$Sn coated Nb cavity cutouts as well as, samples by Focused Ion Beam (FIB) using a Helios 600 FEI instrument at the Materials Research Laboratory (MRL) at the University of Illinois at Urbana-Champaign (UIUC).  A JEOL JEM 2100 LaB$_6$ thermionic gun TEM at MRL/UIUC was used for imaging and nano-area electron diffraction (NED).  The beam size of approximately 100 nm was used to obtain NED patterns.  A Pananalytical/Philips X'pert$^1$ and X'pert$^2$ Material Research Diffractometers at MRL/UIUC was used for XRD.   X'pert$^1$ system is a high-resolution XRD system which is using Cu k-alpha1 radiation with a hybrid mirror (x-ray mirror and a two-reflection Ge monochromator) in the primary optics, and having a PIXcel line detector in the secondary optics.  X'pert$^2$ is a medium-resolution XRD system which is using Cu k-alpha radiation and a parallel beam configuration using a crossed-slit collimator in the primary beam and a parallel plate collimator, flat graphite monochromator and a proportional detector.

A Hitachi HD2300 Dual EDS STEM at Northwestern University (NU) was used for STEM imaging and EDS chemical analysis.  STEM images were taken with a high angle annular dark field detector (HAADF).  EDS maps were taken with incident electron energy of 200 kV.  Frame time was set to 10 s.  Map resolution was set to 512 by 384 pixels.  A Hitachi SU8030 SEM with cold field emission gun equipped with EDS silicon drift detector (SDD) at NU was used for imaging and EDS analytical characterization.  Integrated Oxford AZtec system on FEI Quanta field emission gun SEM at NU was used for EBSD and additional EDS.

\section{Results and Discussion}

\subsection{Structural evaluation}

\begin{figure}[h]
\includegraphics[width=0.5\textwidth]{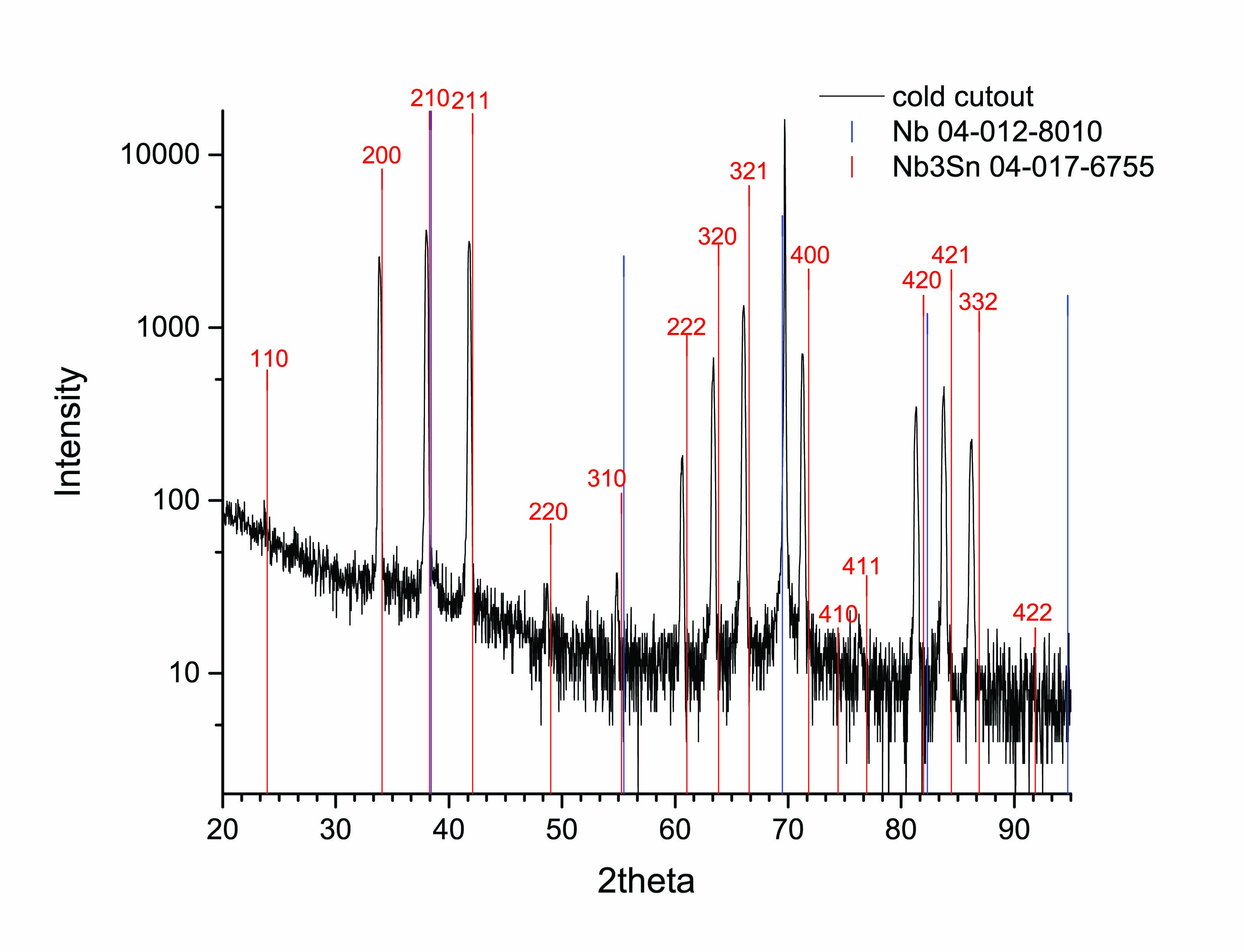}
\centering
\caption{XRD of cold cutout taken with Pananalytical/Philips X'pert$^2$.  Red and blue markers indicated positions of the reference peaks for Nb$_3$Sn and Nb, respectively.}
\label{XRD-cold-Expert2}
\end{figure}

Structural characterization of ERL1-5 cavity cutouts was initiated in order to confirm A15 Nb$_3$Sn structure of the coating and the absence of non-stoichiometric phases, such as Nb$_6$Sn$_5$.  XRD taken with medium resolution X'pert$^2$ diffractometer showed A15 Nb$_3$Sn structure (PDF 04-017-6755) in both, cold and hot cutouts. Fig.$\ref{XRD-cold-Expert2}$ demonstrates data for cold cutout as an example.  Due to the relatively large probing depth of XRD, Nb substrate signal is also present (PDF 04-012-8010).  XRD confirmed the absence of non-A15 phases in the coating.  Lattice constant of Nb$_3$Sn in cold cutout was estimated to be 5.2889 $\mathrm{\AA}$.  Lattice constant of Nb$_3$Sn in hot cutout is 5.2901 $\mathrm{\AA}$.  According to the previous measurements \cite{Godeke}, lattice constants for both cutouts translate into Sn content in the desired range of 24-25 at.\%.  Wavelength-dispersive spectroscopy (WDS) (not shown here) on cold cutout supports this assumption, demonstrating Sn content of 25 at.\%. 

\subsection{Microstructure}

\begin{figure}
\includegraphics[width=0.36\textwidth]{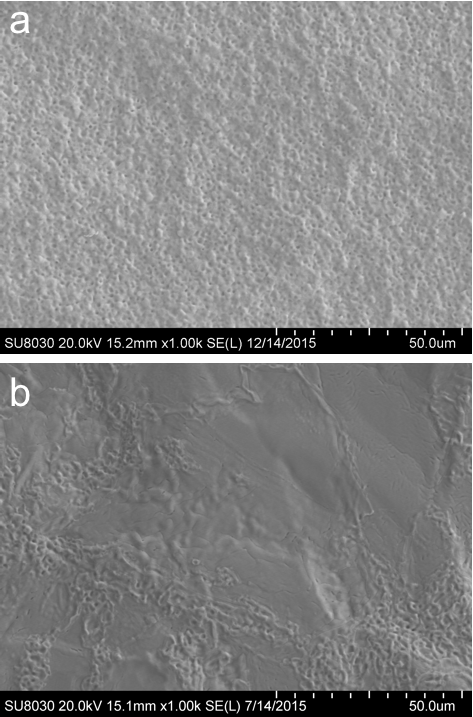}
\centering
\caption{SEM images of (a) cold and (b) hot cutouts.}
\label{SEM-cold-hot}
\end{figure}

SEM images in Fig.$\ref{SEM-cold-hot}$a and b demonstrate the surface of the coating in cold and hot cutouts, respectively.  Cold cutout is uniformly covered with micron-sized Nb$_3$Sn grains.  Hot cutout, on the other hand, shows ``patchy'' regions surrounded by the areas of uniform coating with an appearance similar to the cold cutout surface.  EDS maps of cold cutout taken at the incident accelerating electron voltage of 20 kV are highly uniform and completely featureless (not shown here), demonstrating lateral and in-depth uniformity for Sn and Nb.  Sn content from EDS maps was estimated to 24-25 at.\%.  Since Sn content estimation from EDS agrees well with WDS measurements at the same accelerating voltage, as well as with XRD, we can conclude at an accelerating voltage of 20kV most of EDS signal is coming from within the coating.  

\begin{figure}
\includegraphics[width=0.36\textwidth]{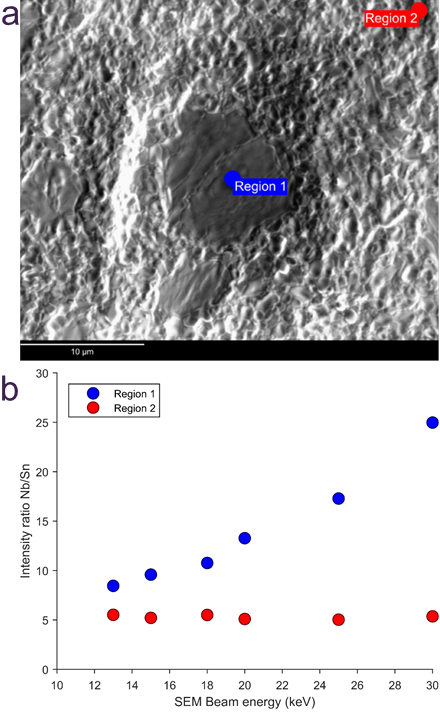}
\centering
\caption{(a) SEM image of Nb$_3$Sn coating taken at Cornell University; (b) Niobium to tin intensity in two regions of the surface vs. SEM incident beam energy.}
\label{Cornell-pict-1}
\end{figure}

\begin{figure*}
\includegraphics[width=0.8\textwidth]{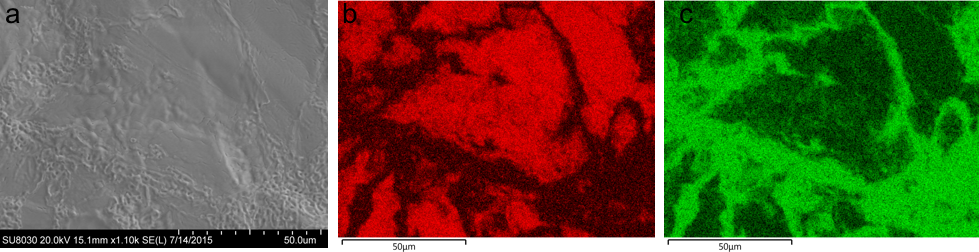}
\centering
\caption{(a) SEM image of hot cutout with corresponding (b) Nb and (c) Sn maps at 20 kV.}
\label{SEM-1}
\end{figure*}

\begin{figure*}
\includegraphics[width=0.7\textwidth]{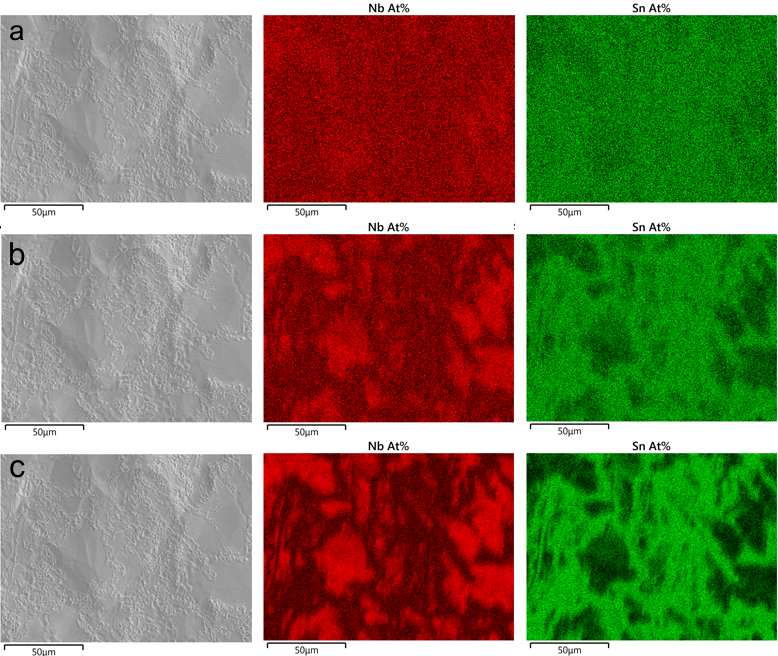}
\centering
\caption{SEM images and EDS maps of Sn and Nb taken from the same spot on hot cuttout at (a) 10KV, (b) 15 kV, and (c) 20 kV.}
\label{SEM-3}
\end{figure*}

\begin{figure}
\includegraphics[width=0.3\textwidth]{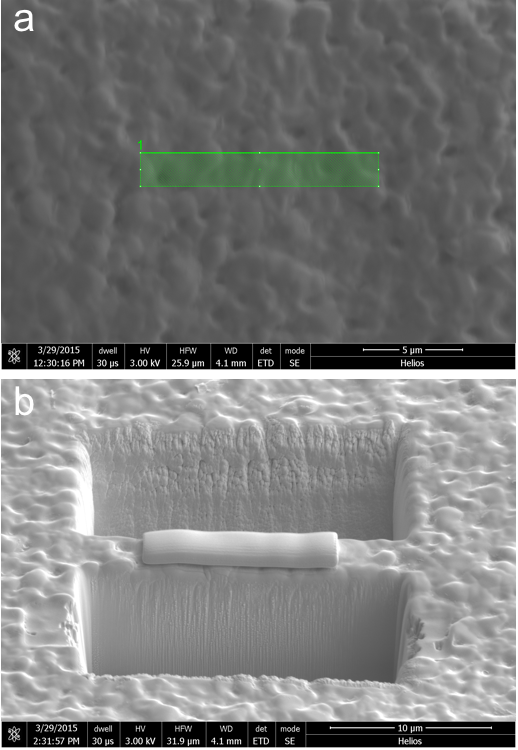}
\centering
\caption{SEM image of Nb$_3$Sn coating in cold cutout (a) prior to FIB sample preparation, (b) with trenches around the cross-sectional cut.}
\label{TUPB056-f2}
\end{figure}

\begin{figure*}
\includegraphics[width=0.8\textwidth]{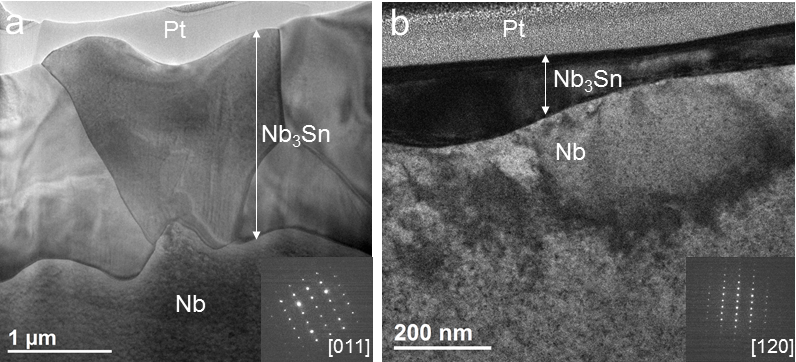}
\centering
\caption{Bright Field image of FIB sample prepared from (a) cold and (b) hot cutouts with corresponding NED patterns.}
\label{TEM-hot-cold}
\end{figure*} 

The appearance of the Nb$_3$Sn coating surface in the hot cutout is very different, compared to the cold one.  Fig.$\ref{SEM-1}$ shows SEM image and corresponding Sn and Nb EDS maps of hot cutout taken at accelerating voltage of 20kV.  Patchy regions show significantly reduced Sn content (darker contrast in Sn map) and elevated Nb concentration (brighter contrast in Nb map).  Reduced Sn signal in EDS map can indicate either significantly reduced Sn concentration in the coating which no longer supports A15 structure or variation in thickness of the coating itself.

By varying the incident electrons accelerating voltage in SEM, one can change probing depth for EDS signal.  Fig.$\ref{SEM-3}$ demonstrates the result of changing the accelerating voltage.  As we increase incident electron voltage, EDS probes a thicker surface layer.  Most of the EDS signal comes from withing hot cutout's coating at the lowest voltage of 10 kV, as can be seen from Fig.$\ref{SEM-3}$a.  Sn concentration taken from the map which was acquired at 10 kV, is close to 24 at.\%.  Much lower Sn and higher Nb content in maps taken at 15 kV and 20 kV (Fig.$\ref{SEM-3}$b and c, respectively) indicates that the EDS signal comes not only from the coating but also from Nb underneath.  Comparing  Fig.$\ref{SEM-3}$a with Fig.$\ref{SEM-3}$b and c, we can conclude that Nb$_3$Sn coating in the patchy regions in hot cutout is much thiner than in cold one.  The hot cutout shows significant variation in Nb$_3$Sn coating thickness across the whole area.

Fig.$\ref{Cornell-pict-1}$a shows SEM picture of the typical patchy region taken at Cornell University.  Fig.$\ref{Cornell-pict-1}$b summarizes the incident voltage varying experiment.  Niobium to tin ratio in the patchy region (Region 1) increases with an incident beam energy, since contribution of niobium from the substrate increases.  Niobium to tin ratio outside the patchy region (Region 2) remains constant with an increase in beam energy, since EDS interaction volume is concentrated within the thickness of the coating. 

Cross-sectional TEM samples had been prepared from cold and hot cutouts to evaluate Nb$_3$Sn coating thickness directly (Fig.$\ref{TUPB056-f2}$).  Fig.$\ref{TEM-hot-cold}$a and b demonstrates TEM bright field (BF) images of cold and hot cutouts, respectively.  Hot cutout FIB sample was taken from the patchy region of the surface which has thinner coating according to SEM studies.  Difference in Nb$_3$Sn coating thickness in cold and hot cutout is obvious when looking at TEM cross-sectional images.  Nb$_3$Sn coating is about 2$\mu$m thick in cold cutout and in the order of only 0.1$\mu$m in the thinnest regions of hot cutout.  More than 10 times difference in Nb$_3$Sn thickness can be noticed comparing cold and hot cutouts.  


\begin{figure}
\includegraphics[width=0.4\textwidth]{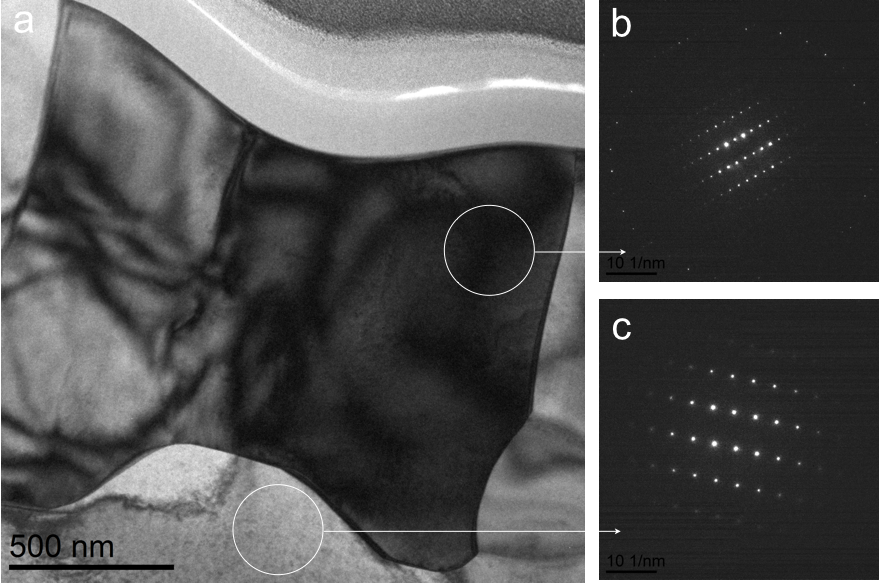}
\centering
\caption{(a) BF image of Nb$_3$Sn coated sample's near-surface; (b) NED taken from within Nb$_3$Sn grain, [120] Nb$_3$Sn zone axis; (c) NED taken below the coating, Nb [113] zone axis.}
\label{TUPB056-f4}
\end{figure}

It is worth mentioning that microstructure of Nb$_3$Sn coating in cold cutout is similar to the coating produced in the flat sample (Fig.$\ref{TUPB056-f4}$) which was coated in the same experimental setup at Cornell University. 

The strong Q-slope observed in ERL1-5 cavity could be explained by the presence of thin  Nb$_3$Sn coating regions in the highly dissipating half-cell.  Coating thickness less than magnetic field penetration depth (about 100nm for  Nb$_3$Sn) leads to superconducting current passing through Nb substrate and  Nb$_3$Sn-Nb interface.  Large surface of one half-cell affected by unacceptably thin coating causes significant dissipation which most probably resulted in Q-slope.

\begin{figure*}
\includegraphics[width=0.9\textwidth]{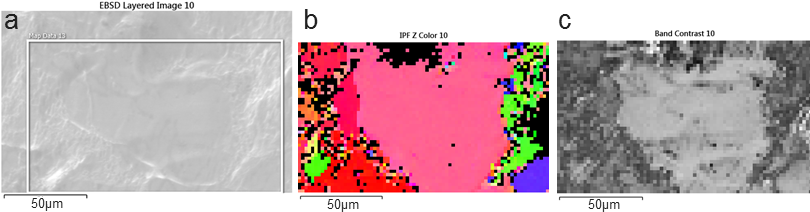}
\centering
\caption{(a) SEM image of the patchy region on hot cutout surface; (b) Inversed Pole Figures (IPF) colors; (c) Band contrast.}
\label{Nb3Sn-EBSD}
\end{figure*}

Microstructure of the coating is also different in cold and hot cutouts.  Coating in cold cutout has approximately micrometer-sized Nb$_3$Sn grains compared to no visible grain boundaries in TEM sample prepared from hot cutout.  As it appears from EBSD taken from one of the patchy regions in the surface of hot cutout (Fig.$\ref{Nb3Sn-EBSD}$), the entire thin (patchy) region is a single Nb$_3$Sn grain of about 100$\mu$m in diameter.  Our conclusions are supported by the analysis of 5 additional TEM samples prepared from cold cutout and 2 additional TEM samples from hot cutout.  Fig.$\ref{TEM-hot-cold}$a shows contrast variation in the center of the largest grain which looks like a presence of some domain.  Appearance of intragrain domains in cold cutout will be explored in the sections below.

\begin{figure}
\includegraphics[width=0.5\textwidth]{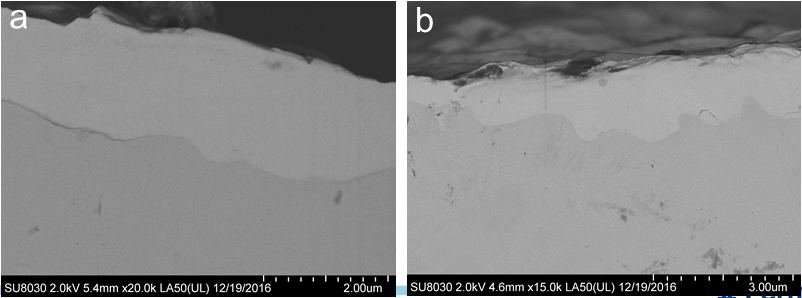}
\centering
\caption{Backscattered images of cross-sectional SEM samples prepared from (a) cold cutout, (b) hot cutout.}
\label{Hot-Cold-CS}
\end{figure}

In order to directly compare thickness of the coating over the area much larger that TEM sample dimensions, cross-sectional samples were prepared from cold and hot cutouts Fig.$\ref{Hot-Cold-CS}$.  Hot sample demonstrates rapidly changing, wavy Nb$_3$Sn/Nb interface which was definitely not induced by Nb substrate.  Cause of such peculiar interface line will be investigated in future studies. 

\subsection{Origin of the thin coating}

\begin{figure}[h]
\includegraphics[width=0.5\textwidth]{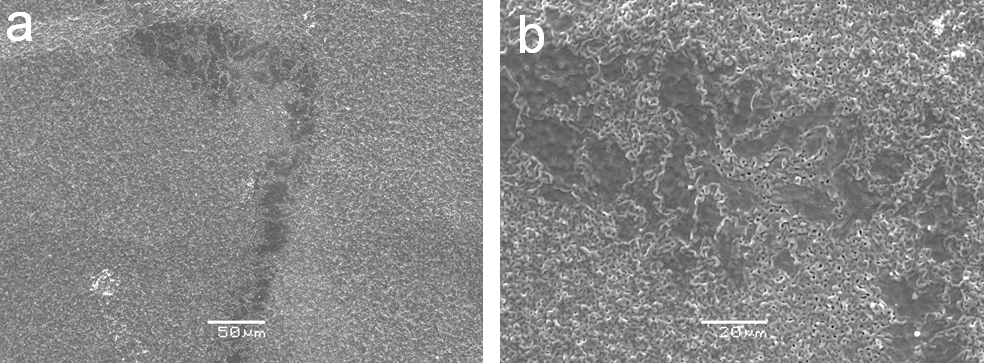}
\centering
\caption{SEM image of the surface of equator spot (a) thin coating region, (b) detailed image of the thin region.}
\label{E1-SEM-good-1}
\end{figure}

The thin Nb$_3$Sn (patchy) regions affect significant area of the hot cutout but were not observed in the cold cutout.  Hot and cold cutouts originate from two different cavity half-cells.  The cause of the patchy regions may be attributed to either Nb substrate or geometry of the deposition process (proximity of the Sn source) or both.  In order to investigate potential causes we explored two cutouts from the equator region of the cavity.  In equator cutouts there is some representative area from both half-cells of the cavity on different sides of the weld.  Most of the surface area in both equator cutouts show uniform Nb$_3$Sn coverage that looks identical to the coating in cold cutout.  However patchy regions of various sizes and shapes were found in random locations in either side of the weld (Fig.$\ref{E1-SEM-good-1}$).  EDS (not shown here) confirmed that the patchy regions in the equator cutouts are similar to the thin coating regions in hot cutout.  The surface area covered by the patchy regions and their size on equator cutouts appear to be higher in shiny half-cell compared to the matte half-cell of Nb$_3$Sn.  Despite the mechanism of patchy regions formation is not clear, their presence in both half-cells of the coated cavity indicate the relevance of the deposition process rather than substrate-related issues.

The presence of the patchy regions may be explained by improper nucleation during the coating process. Siemens researchers showed that failure to properly nucleate the substrate could produce non-uniform regions of size ~100 microns \cite{Hillenbrand-SRF-1980}. They describe two different nucleation methods: 1) anodizing the substrate prior to coating to generate a relatively thick niobium oxide, and, during initial heating, locally raising the temperature of the tin source to transport vapor to the surface while the temperature of the niobium substrate is still relatively low; 2) adding a small amount of SnF2, which has a relatively high vapor pressure at intermediate temperatures, well below the coating temperature \cite{Hillenbrand-SRF-1980}. They explain that in experiments where they only partly anodize substrates, they observe a uniform coating in the anodized areas and incomplete coverage in the non-anodized areas \cite{Hillenbrand-1977}. They suggest that providing a high tin supply pressure to the substrate during early stages of the coating procedure allows a homogeneous nucleation layer to form prior to the niobium oxide dissolving and the tin coming into contact with the niobium surface \cite{Hillenbrand-1977}. 

\begin{figure*}
\includegraphics*[width=0.8\textwidth]{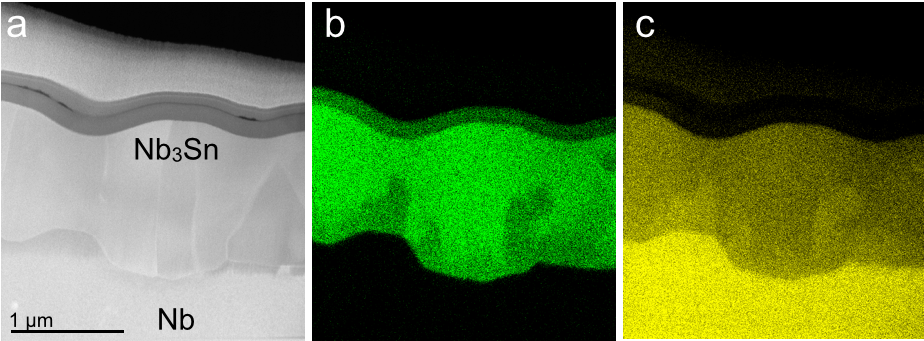}
\centering
\caption{(a) Z-contrast image of the near-surface of the Nb$_3$Sn coated Nb sample; (b) EDS map for Sn; (c) EDS map for Nb.}
\label{TUPB056-f6}
\end{figure*}

\begin{figure*}
\includegraphics*[width=0.8\textwidth]{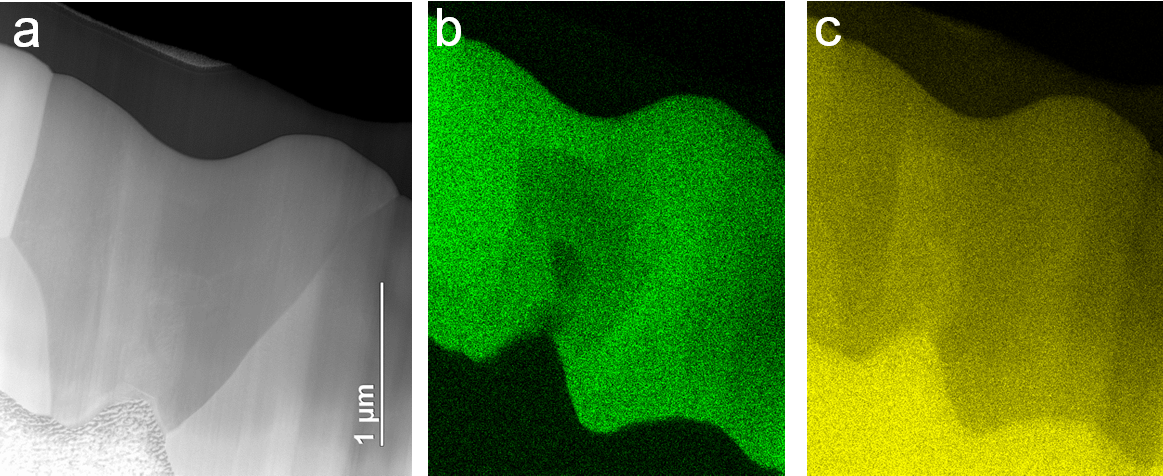}
\centering
\caption{(a) Z-contrast image of the near-surface of cold cutout; (b) EDS map for Sn; (c) EDS map for Nb.}
\label{STEM-EDS-cold}
\end{figure*}

This mechanism of the oxide acting as a buffer between the tin and the niobium until the nucleation layer is formed does not seem to be well explored experimentally. However, the non-uniformities observed in the past appear to be consistent with what is observed in the hot spot cutout. Furthermore, subsequent cavities coated at Cornell, which did not exhibit strong Q-slope or reduced low field Q, received a modified recipe in which the tin source temperature was raised relative to that of the substrate during initial heating, as described in nucleation procedure 1) above. This may have helped to transport tin from the tin source to the substrate earlier, before the nucleation layer generated by the tin halide could fully diffuse into the surface (in the case of the Cornell cavities, SnCl$_2$ was used as a nucleation agent). Currently, this mechanism is largely speculative and much remains to be studied. For example, while Siemens has suggested that the thick oxide (generated by anodizing the substrate) improves Nb$_3$Sn coverage by acting as a buffer between the niobium and the tin, other explanations are quite possible, including a previously proposed catalytic effect of oxygen in the formation of Nb$_3$Sn \cite{Braginski1986}.


\subsection{Chemical characterization}

Regardless of the coating thickness variation in the strongly dissipating half-cell, evaluation of the Nb$_3$Sn quality in the cavity is crucially important for the development of this technology.  Chemical characterization is vital since the superconducting properties of Nb$_3$Sn strongly depend on stoichiometry.  Even a small variation in Sn content has a great impact on superconducting properties of the coating.   The Nb$_3$Sn A15 phase ranges from 18 at.\% to 26 at.\% of Sn \cite{Devantay}.  A critical temperature of approximately 18 K can be reached only for``Sn-rich" composition with approximately 24 at.\% to 26 at.\% of Sn.  In case of lower Sn content in A15 phase, T$_c$ can be as low as 6 K, below traditional Nb, which has a T$_\mathrm{c}$=9.2K.

\begin{figure*}
\includegraphics*[width=0.8\textwidth]{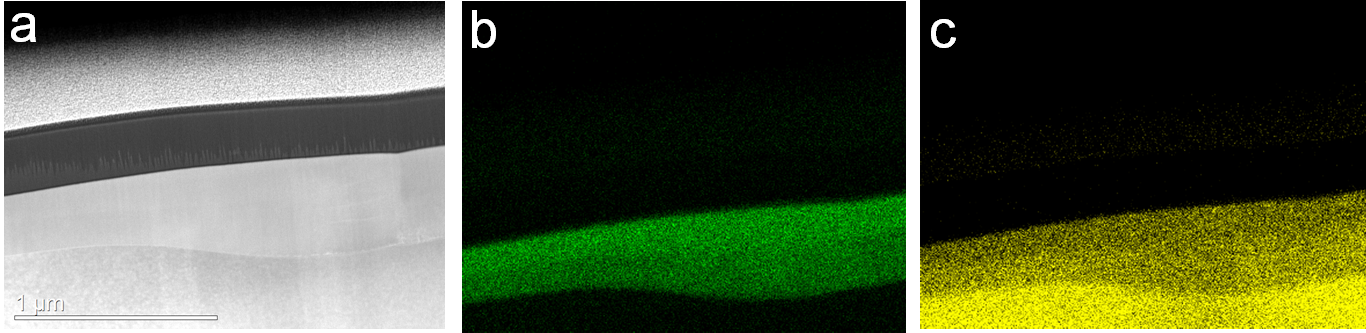}
\centering
\caption{(a) Z-contrast image of the near-surface of hot cutout; (b) EDS map for Sn; (c) EDS map for Nb.}
\label{SEM-EDS-hot-1}
\end{figure*}

The presence of sub-micrometer-sized regions with locally reduced Sn concentration was proposed as a cause of significant reduction of superconducting transition temperature in Nb$_3$Sn coating.  The effect of local compositional inhomogeneity on T$_c$ for A15 compounds was first realized in 80's from specific heat and tunneling studies \cite{ Hellman-1987, Hellman-1986}.

Local composition of Nb$_3$Sn coating in both cutouts and flat sample was explored with STEM/EDS mapping.  Fig.$\ref{TUPB056-f6}$a shows a Z-contrast image of the flat sample's near-surface.  EDS maps of Sn and Nb distribution in the near-surface are represented in Fig.$\ref{TUPB056-f6}$b and c, correspondingly.  Sn map of the near-surface shows several areas with decreased signal intensity in the Nb$_3$Sn coating.  Decreased Sn intensity areas are few hundred nm in size and are located close to the interface between Nb$_3$Sn and Nb.  Decrease in Sn signal intensity in the EDS map can be observed due to either a thiner region of the sample or deficiency of Sn in this particular area.  If the thickness variation would be the case, the Nb map would show the same areas of decreased Nb signal intensity.  However the Nb map shows brighter regions in Nb$_3$Sn coating which indicates higher Nb content.  Also no obvious thickness variation can be observed in corresponding STEM image of the area.  Estimations of the Sn concentration difference between normal and Sn-deficient areas show the difference of 7 to 8 at. \%, regardless of the model used for EDS quantitative evaluation.  Sn-deficiency do not mimic individual Nb$_3$Sn grains.

Sub-micrometer-sized regions of Sn-deficiency, similar to the ones observed in the flat sample's coating, were found in Nb$_3$Sn coating in cold cutout (Fig.$\ref{STEM-EDS-cold}$).  Three samples prepared from the cold cutout and one sample prepared from the flat sample were STEM/EDS measured.  Sn-deficiency regions were found in all of them.  Sn-deficiency regions with likely reduced T$_c$ were found in Nb$_3$Sn-coated cavity for the first time.  The presence of Sn-deficiency regions in the flat sample implies that their formation is inherent to the deposition process rather than the substrate geometry.  Appearance of Sn-deficiency region approximately in the center of the largest grain in cold cutout (Fig.$\ref{STEM-EDS-cold}$) might be explained by deficiency of Sn during grain growth.  Sn from the surface is expected to diffuse along the grain boundaries during the deposition process \cite{Hillenbrand-SRF-1980,Farrell-1974}.  The region in the center of the largest grain is the furthest from the grain boundaries which provide Sn source during the film growth.

Currently the questions of how much and what kind of performance degradation is induced by Sn-deficiency regions in  Nb$_3$Sn cavities remain open.  Since the temperature map of ERL1-5 was taken at 9 MV/m due to rapidly degrading quality factor, the cold cutout doesn't show significant dissipation at such a low gradient.  The Sn-deficiency regions most likely have reduced T$_c$, they are located close the Nb$_3$Sn/Nb interface.  Being few penetration depths away from the cavity surface, Sn-deficiency regions carry much less current comparing to the surface.  However quantitative estimation of Sn-deficiency regions dissipation is not trivial.  The effect of local Sn-deficiency on the the quality factor would need to be evaluated in a cavity free of thin coating regions which most likely cause severe Q-slope.

STEM/EDS characterization of hot cutout is demonstrated in Fig.$\ref{SEM-EDS-hot-1}$.  Two TEM samples prepared from a hot cutout were studied with the STEM/EDS.  It was observed in both TEM samples that the Nb$_3$Sn coating in hot cutout shows reduced Sn signal and brighter Nb signal at the bottom part of the coating.  Observation of lower Sn concentration at the bottom of the coating can be explained by insufficient Sn supply from the grain boundaries during the growth.  Being thin, 100$\mu$m-sized single grain, the patchy region most likely demonstrates Sn-deficiency due to lack of close grain boundaries which transport Sn below the surface.  However signal intensity variations in Sn and Nb maps in hot cutout samples are rather weak.

\begin{figure*}
\includegraphics[width=0.7\textwidth]{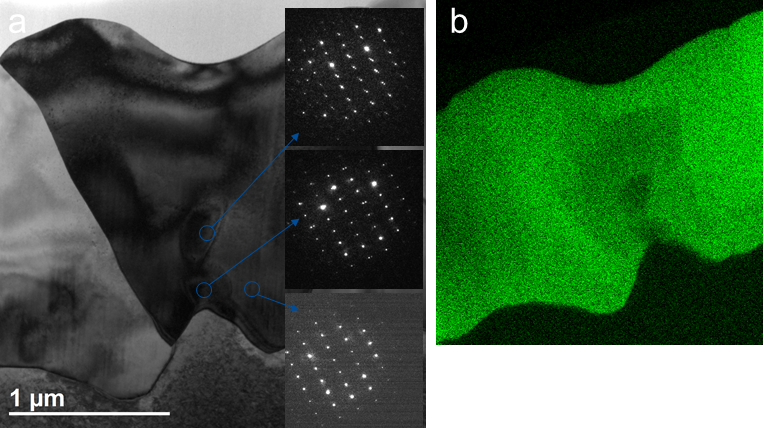}
\centering
\caption{(a)BF TEM image of intragrain domain in cold cutout sample with NED patterns, (b)EDS Sn-map of the same area.}
\label{NED-cold}
\end{figure*}

\subsection{Characterization of Sn-deficiency regions in cold cutout}

Sn-deficiency regions produce contrast variation in the TEM images in some samples (Fig.$\ref{NED-cold}$).  Sn-deficiency regions look like domains in the largest Nb$_3$Sn grains in TEM and TEM BF images.  NED with parallel beam of approximately 100 nm in diameter was used to explore the structure of Sn-deficient domains.  Fig.$\ref{NED-cold}$a, b shows BF TEM image of intragrain domain in cold cutout sample and EDS Sn-map of the same area, respectively.  Corresponding NED show Nb$_3$Sn patterns from intragrain Sn-deficient domain and surrounding area from the same grain.  No change in crystal structure within the grain containing deficiency region was noticed.  However weak additional reflections accompany major diffraction spots inside of Sn-deficiency domain.  Similar additional reflections observed in intragrain domains in all cold cutout TEM samples.

The origin of weak additional reflections in Sn-deficient domain is not yet understood.  The challenge is that most of the observed additional reflections have low intensity under the imaging conditions that were used.  One possible scenario can be realized from theoretical works of previous researchers \cite{Welch-1984}.  Sn-deficiency regions in Nb$_3$Sn A15 structure can be interpreted as accumulation of unit cells where missing Sn atoms are substituted by Nb atoms.  Presence of Nb substitutions on Sn sites in some substantial volume potentially causes various crystal defects which were described in theoretical calculations.  Possibility of anti-phase domains (like in Au-Cu alloys) can also be considered \cite{ MARCINKOWSKI1963, Zhu-1982}.

\begin{figure}
\includegraphics[width=0.46\textwidth]{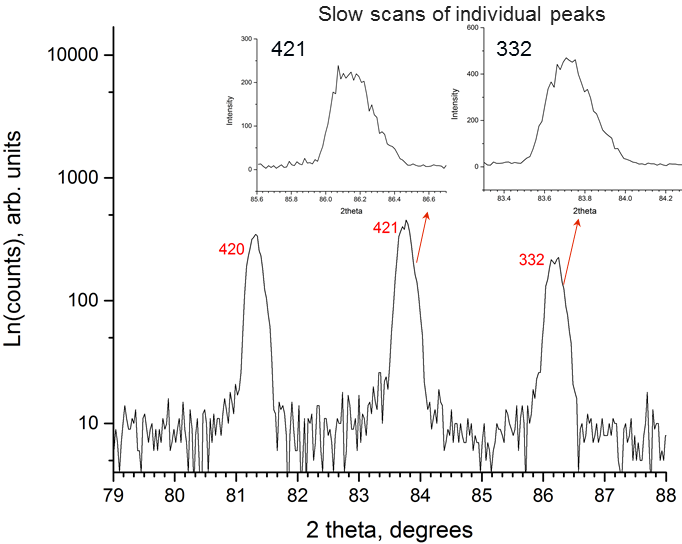}
\centering
\caption{High resolution slow XRD scan of cold cutout.}
\label{XRD-cold}
\end{figure}

Effect of compositional variation in Sn-deficiency regions on global properties of the film was explored with high resolution XRD performed with X'pert$^1$ diffractometer.  One would expect reduction of the lattice constant with lower Sn concentration.  Since A15 Nb$_3$Sn structure was already confirmed in cold cutout, shape of diffraction peaks was investigated with higher resolution.  Fig.$\ref{XRD-cold}$ shows several diffraction peaks which were slowly scanned with the best possible resolution and count rate.  Every Nb$_3$Sn diffraction peak in cold cutout shows some asymmetry in the peak shape.  Asymmetry extends toward higher diffraction angles.  According to Bragg's law diffraction peaks at higher angles indicate smaller d-spacings in diffracting material which, for cubic structures, translates into the smaller lattice constant.  Previous experiments \cite{Godeke} show decrease in T$_\mathrm{c}$ of Nb$_3$Sn with reduced lattice constant which was induced by lower Sn concentration.  Consistent asymmetry of diffraction peaks shape indicates a spread in d-spacings in diffracting material which can be interpreted as presence of A15 Nb$_3$Sn with a few slightly different lattice constants.  Smaller lattice constants are most certainly originating from Sn-deficiency regions observed by STEM/EDS.  Presence of a spread in lattice constants was previously reported in Nb$_3$Sn coating prepared by the same method \cite{Becker}.

Identical experimental conditions were used to explore diffraction peak shapes in hot cutout.  However no asymmetry of the peaks was found in hot cutout which implies no spread in lattice constants values within available resolution. 

\section{Conclusion}

For the first time, extended material characterization of original Nb$_3$Sn-coated cavity cutouts with known dissipation profiles demonstrated Nb$_3$Sn features responsible for degraded SRF performance.  Comparison of hot cavity cutout with elevated surface resistance which causes significant heating, and cold cutout shows drastic difference in the Nb$_3$Sn coating thickness.  Nb$_3$Sn coating thickness under 100nm in hot cutout was observed.  We suspect that insufficient coating thickness in hot cutout enables penetration of superconducting current into underlaying Nb and Nb$_3$Sn-Nb interface causing rapid Q-drop.  In hot cutout, Nb$_3$Sn grain diameter in thin (patchy) regions approaches 100$\mu$m.  Formation of large, thin grains can be explained by insufficient Sn transport due to a low density of grain boundaries during the deposition process \cite{Hillenbrand-1977,Farrell-1974}.

Sub-micrometer-sized local composition variation regions were found in non-dissipating (at measurement conditions) cold cutout.  Presence of Sn-deficiency regions in cold cutout was directly observed with TEM imaging and STEM/EDS analysis in multiple TEM samples.  Effect of the lattice constant due to Sn deficiency was found with high resolution XRD.  Sn-deficiency regions were found only in cold cutout.  Our studies suggest that Sn-deficiency regions with suppressed T$_\mathrm{c}$ can certainly affect the performance of Nb$_3$Sn-coated cavities.  Extended material characterization of coupons from a Nb$_3$Sn-coated cavity without drastic Q-slope is needed to evaluate the severity of performance degradation due to the local Sn-deficiency.  

Structural and analytical characterization of Nb$_3$Sn coating in combination with T-mapping during the RF test provides figures of merit for Nb$_3$Sn material quality.  Despite additional studies needed to fully understand the structure of Sn-deficiency regions, this work established the first example of the combined material and RF Nb$_3$Sn-coated cavity characterization.  For Nb$_3$Sn program at Fermilab, this study provided a trial for the characterization tool set which will be routinely used for the coatings produced at Fermilab.

\section*{Acknowledgement}

This work was carried out in part in the Frederick Seitz Material Research Laboratory Central Research Facilities, University of Illinois.  The authors would like to thank all the stuff scientists who work at Center for Microanalysis of Materials in Frederick Seitz Material Research Laboratory.  This work made use of the EPIC, Keck-II, and/or SPID facility(ies) of Northwestern University’s NUANCE Center, which has received support from the Soft and Hybrid Nanotechnology Experimental (SHyNE) Resource (NSF ECCS-1542205); the MRSEC program (NSF DMR-1121262) at the Materials Research Center; the International Institute for Nanotechnology (IIN); the Keck Foundation; and the State of Illinois, through the IIN.

Work performed at Cornell was primarily supported by U.S. DOE award DE-SC0008431, and in part by the U.S. National Science Foundation under Award PHY-1549132, the Center for Bright Beams. Select results presented in this work made use of the Cornell Center for Materials Research Shared Facilities which are supported through the NSF MRSEC program (DMR-1120296).

\section*{References}
 
\bibliography{library}
\bibliographystyle{unsrt}
 
\end{document}